\documentclass[a4paper,11pt]{article}
\pdfoutput=1 % if your are submitting a pdflatex (i.e. if you have
\usepackage{jheppub} % for details on the use of the package, please
\usepackage{amsmath}
\usepackage[T1]{fontenc} % if needed
\usepackage{graphicx}
\usepackage{subfig}

\title{Thermodynamics and evaporation of perfect fluid dark matter black hole in phantom background}% Force line breaks with \\
\author[a]{Xiao Liang,}
\author[a,b,d]{Ya-Peng Hu,}
\author[a]{Chen-Hao Wu,}
\author[1,a,c]{Yu-Sen An\note{Corresponding author.}}

\affiliation[a]{College of Physics, Nanjing University of Aeronautics and Astronautics, Nanjing, 210016, China}

\affiliation[b]{MIIT Key Laboratory of Aerospace Information Materials and Physics,  Nanjing University of Aeronautics and Astronautics, Nanjing, 210016, China}

\affiliation[c]{Department of Physics and Center for Field Theory and Particle Physics, Fudan University, Shanghai 200433, China}

\affiliation[d]{Center for Gravitation and Cosmology, College of Physical Science and Technology, Yangzhou University, Yangzhou, 225009, China}

\emailAdd{xliang@nuaa.edu.cn,huyp@nuaa.edu.cn,chenhao\_wu@nuaa.edu.cn,
anyusen@nuaa.edu.cn}

\abstract
{We present a novel interpretation of the thermodynamics of perfect fluid dark matter (PFDM) black hole based on Misner-Sharp energy, and then investigate its evaporation behavior. We find that the ratio between dark sector initial density and black hole horizon radius significantly influences black hole evaporation behaviors. We demonstrate that the presence of the dark sector can significantly extend the lifetime of a black hole which is similar to the Reissner-Nordstrom case. Our work reformulates the thermodynamics of PFDM black holes and points out the existence of long-lived black holes in the presence of the dark sector.}

\begin{document} 
\maketitle
\flushbottom
\section{Introduction}

Several astrophysical observations provide evidence that supermassive black holes, surrounded by massive dark matter halos,  reside in the centers of giant elliptical and spiral galaxies \cite{Rubin:1980zd,EventHorizonTelescope:2019dse,EventHorizonTelescope:2019ggy} (see also \cite{VCRubin,MPersic,GBertone} for related studies). Therefore, when investigating black holes in actual astrophysical scenarios, it is imperative to account for the effect of dark matter \cite{Kiselev:2002dx}. Meanwhile, the results of Supernova searches\cite{SupernovaCosmologyProject:1998vns, SupernovaSearchTeam:1998fmf}, WMAP\cite{WMAP:2003elm}, and Planck\cite{Planck:2015fie, Planck:2018vyg} reveal the accelerating expansion of the universe. To explain this, physicists construct a brand-new physical model: the universe contains something with negative pressure called dark energy\cite{Copeland:2006wr, Peebles:2002gy}. Among various models of dark energy \cite{Bamba:2012cp}, one well-known dark energy model is the phantom energy \cite{Caldwell:1999ew, Caldwell:2003vq}. Thus investigating black hole solutions in the presence of dark matter and dark energy is a fascinating scientific endeavor. The simplest solution is a spherically symmetric black hole immersed in a perfect fluid dark matter (PFDM) within a phantom dark energy background, as first derived in Ref.\cite{Li:2012zx}. This reference 
 discussed the stability of the black hole and various other observational phenomena. The PFDM black hole solution also has been followed by extensive discussions in Ref.\cite{Haroon:2018ryd,Konoplya:2019sns,Jusufi:2019nrn,Narzilloev:2020qtd,Rizwan:2018rgs,Stuchlik:2019uvf,Schee:2019gki,Hendi:2020zyw,Shaymatov:2020wtj,Xu:2016ylr,Cao:2021dcq}. Despite above progress made, certain crucial thermodynamic properties, such as the evaporation of this PFDM black hole, still remain poorly explored.

Hawking radiation established the connection between the four laws of black hole mechanics and the four laws of thermodynamics. Researchers have thus begun treating black holes as thermodynamic systems, and developing black hole thermodynamics \cite{Hawking:1976de,Hawking:1982dh}. The Hawking radiation indicates that a black hole can radiate energy outside and evaporate away. Moreover, the detailed Hawking radiation spectrum allows the calculation of the rate of particle emission and estimation of the lifetime of the black hole. In the framework of quantum field theory in curved space-time, Hawking pioneered a discussion of black hole evaporation and demonstrated that black holes evaporate at an increasing rate until nothing remains \cite{Hawking:1974rv,Hawking:1975vcx}. In 1976, Page performed numerical calculations about the lifetime of a black hole in four-dimensional asymptotically flat spacetime. He found that there is a simple relation between a black hole's lifetime and its initial mass $t \sim M_0^3$ \cite{Page:1976df},  and then extended these discussions to rotating black hole case \cite{Page:1976ki}. In addition, by applying Hawking radiation and the Schwinger effect, the charged black hole will keep losing its charge and mass. It is worth noting that this evaporation process is intriguing  because the charge-over-mass ratio can increase rapidly, and hence the black hole quickly gets close to the extremal black hole. This mechanism will largely extend the lifetime of Reissner-Nordstrom(RN) black holes\cite{Hiscock:1990ex,Xu:2019wak}. Based on Hawking radiation, the study on black hole evaporation in asymptotically flat space-time has proven to be a fruitful research area. For more evaporation processes in various black hole backgrounds see Ref.\cite{Wu:2021zyl,Hou:2020yni,Xu:2020xsl}. Since the thermodynamics and evaporation process are of great importance to the study of black holes, it would be interesting also to investigate them in PFDM black hole background.

In this paper, we investigate the thermodynamics and evaporation of PFDM black holes. In section \ref{sec2}, we give a review of the PFDM black hole solution. In section \ref{thermo}, we investigate the thermodynamic properties of this black hole. And finally, in section \ref{evop}, we discuss the evaporation behavior of this black hole. Throughout this paper, the Misner Sharp energy is an important quantity. In appendix \ref{secdil}, we discuss the relation between Misner Sharp energy and black hole total energy in RN and PFDM black holes.

\section{Review of the model} \label{sec2}
For perfect fluid dark matter black hole background, there exists a non-homogeneous phantom field that is minimally coupled to gravity and a dark matter field that weakly interacts with the phantom field. The action reads \cite{Li:2012zx}
\begin{equation}
    S=\int d^{4}x \sqrt{-g}[\frac{R}{2\kappa^{2}}+\frac{1}{2}g^{\mu\nu} \partial_{\mu}\Phi \partial_{\nu}\Phi-V(\Phi)+\mathcal{L}_{m}+\mathcal{L}_{I}]\label{act}
\end{equation}
where $\kappa^{2}=8\pi G$, $\mathcal{L}_{m}$ is massive dark matter in the galaxy and $\mathcal{L}_{I}$ is the interaction between dark matter and phantom field. Because the coupling between dark matter and phantom field is very small, we will ignore $\mathcal{L}_{I}$ throughout the paper.

First of all, considering the static and spherical symmetric case, the metric ansatz is 
\begin{equation}
    ds^{2}=-e^{\nu(r)}dt^{2}+e^{\mu(r)} dr^{2}+r^{2}(d\theta^{2}+\sin^{2}\theta d\phi^{2}).
\end{equation}
Also, the ansatz of the phantom field is given by $\Phi=\Phi(r)$.  Varying matter action in (\ref{act}) with respect to the metric field, the stress-energy tensor $T_\nu^\mu$ can be obtained as follow, where the dark matter ($\mathcal{L}_{m}$) is treated as a perfect fluid:
\begin{equation}
    T^{t}_{t}=-\rho=\frac{1}{2}e^{-\mu}\Phi'^{2}-V(\Phi)-\rho_{DM}\label{ttt}
\end{equation}
\begin{equation}
    T^{r}_{r}=p_{r}=-\frac{1}{2}e^{-\mu}\Phi'^{2}-V(\Phi)\label{trr}
\end{equation}
\begin{equation}
T^{\theta}_{\theta}=p_\theta=T^{\phi}_{\phi}=p_\phi=\frac{1}{2}e^{-\mu}\Phi'^{2}-V(\Phi)
\end{equation}
 the prime $'$ is the derivative with respect to $r$ and the dark matter stress-energy tensor is given by $T_{\nu}^{\mu (DM)}=\mathrm{diag}(-\rho_{DM},0,0,0)$.

From the above metric ansatz,  Einstein equations can be written as:
\begin{equation}\label{ett}
e^{-\mu}(\frac{1}{r^{2}}-\frac{\mu'}{r})-\frac{1}{r^{2}}=8\pi G T^{t}_{t}
\end{equation}
\begin{equation}\label{err}
e^{-\mu}(\frac{1}{r^{2}}+\frac{\nu'}{r})-\frac{1}{r^{2}}=8\pi G T^{r}_{r}
\end{equation}
\begin{equation} \label{ethth}
    \frac{e^{-\mu}}{2}(\nu''+\frac{\nu'^{2}}{2}+\frac{\nu'-\mu'}{r}-\frac{\nu' \mu'}{2})=8\pi G T^{\theta}_{\theta}=8\pi G T^{\phi}_{\phi}
\end{equation}

As we are interested in finding solutions with the condition $\mu(r)=-\nu(r)$, which imposes the constraint $T^t_t=T^r_r$ from Eq.(\ref{ett}) and Eq.(\ref{err}), the mass density of WIMPs $\rho_{DM}=e^{\nu}\Phi'^2$ can be obtained from (\ref{ttt}) and (\ref{trr}). It is important to note that such a solution can be obtained thanks to the negative kinetic energy term of the phantom field, which is different from the ordinary scalar field like quintessence.

Furthermore, if the stress-energy tensor obeys the relation $T^{\theta}_{\theta}=T^{\phi}_{\phi}=-\frac{1}{2}T^{t}_{t}$ \footnote{Generally, $T^{\theta}_{\theta}=\gamma T^{t}_{t}$, where $\gamma=0,+1,-1$ correspond to Schwarzschild, Schwarzschild dS/AdS, RN black hole respectively, here we are only interested in $\gamma=-\frac{1}{2}$ case.}, by combining Eq. (\ref{ett}) and (\ref{ethth}) together,  we find that 
\begin{equation}\label{eq1}  e^{\nu}r^{2}\nu''+e^{\nu}r^{2}\nu'^{2}+3e^{\nu}r \nu'+e^{\nu}-1=0
\end{equation}
Using variable $U=1-e^{\nu}$, the Eq.(\ref{eq1}) can be written as simple form 
\begin{equation} \label{eqsimple}
    r^{2}U''+3rU'+U=0.
\end{equation}
Eq.(\ref{eqsimple}) can be easily solved as $U=\frac{2M}{r}+\frac{\lambda}{r} \log \frac{r}{\lambda}$ , thus we find a non-trivial black hole solution which reads~\cite{Li:2012zx}
\begin{equation} \label{metr}
       ds^{2}=-(1-\frac{2M}{r}-\frac{\lambda}{r} \log \frac{r}{\lambda})\mathrm{d}t^{2}+(1-\frac{2M}{r}-\frac{\lambda}{r} \log \frac{r}{\lambda})^{-1}\mathrm{d}r^{2}+r^{2}(\mathrm{d}\theta^{2}+\sin^{2}\theta \mathrm{d}\phi^{2})
\end{equation}
where $M$ and $\lambda$ represent the integration constants and we use convention $\lambda>0$. Note that the relation between $T^{\theta}_{\theta}$ and $T^{t}_{t}$ gives constraint of the potential $V(\Phi)$, we can easily solve $V(\Phi)=\frac{1}{6}e^{-\mu}\Phi'^{2}$, thus we find $\rho=\frac{2}{3}e^{-\mu}\Phi'^{2}>0$.

According to Einstein equation (\ref{ett}), the total energy density and dark matter energy density can be obtained as follows
\begin{equation}\label{dark}
   \rho =\frac{\lambda}{8\pi r^{3}}\quad \rho_{DM}=\frac{3 \lambda}{16 \pi r^{3}},
\end{equation}
the parameter $\lambda$ is usually referred to as `dark matter' density in previous work \cite{Hendi:2020zyw,Xu:2016ylr,Cao:2021dcq}. However here, according to Eq.(\ref{dark}) the parameter $\lambda$ actually relates to the total energy density (dark matter combined with the phantom dark energy), where the dark matter energy density and phantom dark energy density differ by a proportional constant. Therefore, strictly speaking, we should refer to $\lambda$ as `dark sector' density based on the above consideration, and we will use dark sector instead of dark matter in the thereafter work.

%\textcolor{red}{For the previous metric function
%\begin{equation}
%ds^{2}=-(1-\frac{2M}{r}+\frac{\lambda}{r} \log \frac{r}{\lambda})dt^{2}+(1-\frac{2M}{r}+\frac{\lambda}{r} \log \frac{r}{\lambda})^{-1}dr^{2}+r^{2}(d\theta^{2}+sin^{2}\theta d\phi^{2})
%\end{equation}
%The energy density $\rho=-\frac{\lambda}{8\pi r^{3}}<0$. It seems wrong to choose this metric!. }
\section{Thermodynamics of PFDM black hole } \label{thermo}
\begin{figure}[!ht]
	\centering
	\includegraphics[width=0.8\textwidth,keepaspectratio]{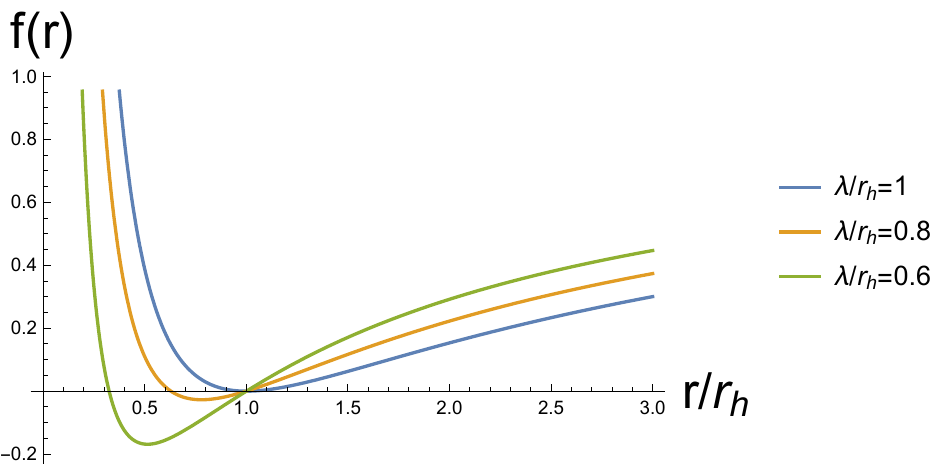}
 \caption{Metric function with different $\lambda/r_{h}=1,0.8,0.6$, we see when increasing $\lambda$, the inner and outer horizon approach each other and finally merge together,  $\lambda/r_{h}=1$ is the extremal black hole case.} \label{3}
\end{figure}

In the preceding section, it has been demonstrated that the metric of the black hole is formulated as Eq.(\ref{metr}).
The blackening factor can also be rewritten in terms of outer horizon radius $r_{h}$ which is the largest root of $f(r)=0$ as 
\begin{equation}
    f(r)=1-\frac{2M}{r}-\frac{\lambda}{r} \log \frac{r}{\lambda}=1-\frac{r_{h}}{r}-\frac{\lambda}{r} \log \frac{r}{r_{h}}
\end{equation}
From $f(r)$, the temperature and entropy of this black hole can be easily computed as 
\begin{equation}
    T=\frac{f'(r_{h})}{4\pi}=\frac{r_{h}-\lambda}{4\pi r_{h}^{2}}, \quad S=\frac{A}{4}=\pi r_{h}^{2}.
\end{equation}
From the above thermodynamic quantities, as the temperature should be non-negative, we find $r_{h}\geqslant \lambda$, $\lambda=r_h$ corresponds to the extremal black hole case. \footnote{The $\lambda>r_{h}$ case is forbidden to avoid the naked singularity.}  
It is also easy to show that for $r_{h}>\lambda$, PFDM black holes have two horizons. When  $\lambda/r_h=1$, two horizons coincide with each other. To be more explicit, we plot the blackening factor $f(r)$ for different $\lambda/r_{h}$ in Fig.\ref{3}.

Next, we discuss the energy definition and thermodynamic relations of this system. Firstly, we can naively calculate the Komar mass at infinity. The Komar mass is defined by 
\begin{equation}
      E_{R}=\frac{1}{4\pi} \int_{\partial \Sigma} d^{2}x \sqrt{\gamma^{(2)} }n_{\mu}\sigma_{\nu} \nabla^{\mu} K^{\nu}
\end{equation}
where $K^{\mu}=(1,0,0,0)$ is the asymptotic Killing vector, $\partial \Sigma$ is the co-dimension two sphere at infinity and $n_{\mu}$, $\sigma_{\nu}$ is the time-like and space-like normal vector of $\partial \Sigma$ respectively
\begin{equation}
    n_{0}=-(1-\frac{2M}{r}-\frac{\lambda}{r} \log \frac{r}{\lambda})^{1/2}
\end{equation}
\begin{equation}
    \sigma_{1}=(1-\frac{2M}{r}-\frac{\lambda}{r} \log \frac{r}{\lambda})^{-1/2}.
\end{equation}
So 
\begin{equation}
n_{\mu} \sigma_{\nu} \nabla^{\mu} K^{\nu}=-\nabla^{0} K^{1}=-g^{00} \Gamma^{1}_{00} K^{0}=\frac{1}{2} (\frac{2M}{r^{2}}+\frac{\lambda}{r^{2}} \log \frac{r}{\lambda}-\frac{\lambda}{r^{2}})
\end{equation}
where we use the relation
\begin{equation}
    \Gamma^{1}_{00}=-\frac{1}{2}g^{11}g_{00,1}=\frac{1}{2}(1-\frac{2M}{r}-\frac{\lambda}{r}\log \frac{r}{\lambda})\partial_{r}[1-\frac{2M}{r}-\frac{\lambda}{r} \log \frac{r}{\lambda}].
\end{equation}
So the Komar mass is 
\begin{equation}
    E_{R}=M+\frac{\lambda}{2} \log \frac{r_{\infty}}{\lambda}-\frac{1}{2}\lambda=\frac{1}{2}r_{h}+\frac{\lambda}{2} \log \frac{r_{\infty}}{r_{h}}-\frac{1}{2}\lambda.
\end{equation}
The Komar mass diverges, this divergence is related to the diverging dark sector potential difference. We will show this in the following, if we take this Komar mass as the total energy of spacetime, we can find that in this case the thermodynamic first law is written as 
\begin{equation}
    dE_{R}=TdS+(\frac{1}{2}\log \frac{r_{\infty}}{r_{h}}-\frac{1}{2})d\lambda.
\end{equation}
Analogous to RN black hole case \cite{Kubiznak:2012wp,Kubiznak:2016qmn}, the chemical potential of the dark sector should be identified as 
\begin{equation}
    \mu=\frac{1}{2}\log \frac{r_{\infty}}{r_{h}}-\frac{1}{2}
\end{equation}
which is interpreted as the dark sector potential difference between infinity and horizon.\footnote{Here there is a mismatch $1/2$ which seems unnatural, while below we will see that this mismatch will disappear if we reinterpret the thermodynamics using Misner-Sharp energy at $r=r_{halo}$.} We see as a result of the divergent dark sector potential difference between infinity and the horizon, both the Komar mass and chemical potential exhibits a logarithmic divergence.  This divergence is non-physical because of the unphysical assumption that the dark sector exists all along to spatial infinity. Since the dark sector potential here is of logarithmic confinement type, its existence must be confined within a certain region $r<r_{halo}$ \footnote{Generally, $r_{halo}$ may change through the evaporation process. While here, we assume $r_{halo}$ to be a constant in this toy model instead of a variable}. In other words, the dark sector forms a halo around the black hole, inside the $r=r_{halo}$ sphere. For the region $r>r_{halo}$, the stress tensor will be the vacuum stress tensor as we only focus on the asymptotically flat case. It is the same spacetime structure as in Ref.\cite{Li:2012zx}, while here we give an argument from the thermodynamic point of view.

Because all the matter field is inside $r_{halo}$, it is desirable to use the Misner-Sharp energy inside $r_{halo}$ to denote the energy of spacetime. 
Note that Misner-Sharp energy inside one surface at $r$ is defined by \cite{Misner:1964je,Abdusattar:2021wfv,Hu:2015xva,Zhang:2014goa}
\begin{equation}
M_{MS}=\frac{r}{2}(1-h^{ab}\partial_{a} r\partial_{b} r).
\end{equation}
Thus the Misner-Sharp mass inside $r_{halo}$ is 
\begin{equation}\label{misner}
    M_{MS}|_{r_{halo}}=\frac{r_{halo}}{2}(1-(1-\frac{2M}{r_{halo}}-\frac{\lambda}{r_{halo}} \log \frac{r_{halo}}{\lambda}))=\frac{1}{2}r_{h}+\frac{\lambda}{2}\log \frac{r_{halo}}{r_{h}}
\end{equation}
where we use the relation $M=\frac{1}{2}r_{h}-\frac{\lambda}{2}\log \frac{r_{h}}{\lambda}$. 
We can also use this Misner-Sharp energy to replace the parameter $M$ by relation $M=M_{MS}|_{r_{halo}}-\frac{\lambda}{2}\log\frac{r_{halo}}{\lambda}$ in the metric function which gives
\begin{equation}\label{metric}
f(r)=1-\frac{2M_{MS}|_{r_{halo}}}{r}+\frac{\lambda}{r} \log \frac{r_{halo}}{r}.
\end{equation}
Beyond $r_{halo}$, as the stress-energy tensor is the vacuum stress tensor, the metric behaves as the Schwarzschild spacetime, and the metric should be 
\begin{equation}
ds^{2}=-(1-\frac{2M_{MS}|_{r_{halo}}}{r})dt^{2}+(1-\frac{2M_{MS}|_{r_{halo}}}{r})^{-1} dr^{2}+r^{2}d\Omega^{2},
\end{equation}
where $r>r_{halo}$. Note that in this case, the Komar mass at infinity is $M_{MS}$ and free from divergence, thus it is natural to identify this Misner-Sharp energy $M_{MS}$ inside $r_{halo}$ as the total energy $E$ of the spacetime.

And then we can find the corresponding thermodynamic first law \footnote{This relation holds for an observer at asymptotic infinity, first term at RHS $TdS=\delta Q$ term is the heat transport term while the second term is the work term. For thermal radiation, the evolution of heat transport term follows Stephan-Boltzman law.} 
\begin{equation}
dE=TdS+\mu d\lambda=TdS+\frac{1}{2} \log \frac{r_{halo}}{r_{h}} d\lambda,
\end{equation}
where $E=M_{MS}$ in Eq.(\ref{misner}) and analogous to RN case, $\mu=\frac{1}{2} \log \frac{r_{halo}}{r_{h}}$ is identified as the dark sector chemical potential, which is the potential difference between $r=r_{halo}$ and $r=r_{h}$.

Note that our thermodynamic relation is different from previous work \cite{Xu:2016ylr,Cao:2021dcq}. In previous work, there is a finite
ambiguity in the thermodynamic first law which is related to the ambiguity where we choose as the
reference point of dark sector potential. Also, they use mass parameter M as the definition of energy, but
as we show above, it is no longer the energy of space-time when there is a dark sector halo. 

\section{Evaporation process of black hole} \label{evop}

In the previous section, we have clarified what the thermodynamics of the PFDM black hole looks like. As a black hole behaves like a thermal system, it can radiate away energy and evaporate. Since the evaporation process is an important part of black hole thermodynamics, we will discuss the evaporation behavior of PFDM black holes in this section. 

First note again that for black hole temperature be positive, the horizon radius $r_{h}$ should obey the relation $r_{h}>\lambda$. Like RN black hole, there is an extremal black hole configuration where $r_{h}=\lambda$, which has zero temperature. 

Depending on the evolution behavior of dark sector density $\lambda$ during evaporation, the evaporation behavior can be very different. 

If dark sector density $\lambda$ does not decay, $r_{h}$ will decrease due to the black-body radiation and finally reach the extremal black hole. It can be shown that the time to reach an extremal black hole diverges which reflects the third law of black hole thermodynamics.

While more physically, because the presence of large field strength can turn the virtual particles into real particle anti-particle pairs, the quantum vacuum can have pair productions of the dark sector in the presence of large field strength like the Schwinger effect in quantum electrodynamics (QED) \cite{Gibbons:1975kk,Schwinger:1951nm}. The dark sector density $\lambda$ can decay and black hole evaporation behavior will be different because of this dark sector decay. Although we do not know the detailed physical mechanisms for the evolution of $\lambda$ as it will depend on the microscopic model of dark sector, we can still assume that the pair production rate of dark sector particles is similar to pair production rate in QED.\footnote{At least for some $U(1)$ model which has similar action as QED, the pair production should be similar because the computation of Euler-Heisenberg effective action is similar.}  We leave the detailed calculation of the Schwinger pair production mechanism in the microscopic  model to future work. Readers who are interested in it can also consult paper \cite{Suganuma:1991ha,Nayak:2005pf} for Schwinger effect in QCD, where the potential is also of confining type. The final result of quark-antiquark pair production in \cite{Suganuma:1991ha,Nayak:2005pf} is similar to the QED case.

%\begin{itemize}
  %  \item For fast evaporation rate of $\lambda$ \footnote{Fast or slow decay may depend on the critical field strength, which depends on specific dark matter model} , it will quickly drop to 0, black hole will evaporate away quickly like Schwarzschild black hole.
   % \item For slow evaporation of $\lambda$, then black hole radius $r_{h}$ will first drops close to $\lambda$ due to thermal radiation, the evaporation process will be very slow due to the small temperature, thus the existence of the dark matter will make the black hole long-lived.
%\end{itemize}

%Although we do not know the detailed physical mechanisms for the evolution of $\lambda$ , as this will depend on the specific microscopic model of dark matter \footnote{Here the dark matter stress tensor is approximated as the perfect fluid which hides all the microscopic details. }. We list all types of influence the dark matter can have on the black hole evaporation. We leave the investigation of possible Schwinger pair production like effect in microscopic dark matter model to future work. Moreover it is also interesting to see if we could use the observation of black hole lifetime and evaporating behavior to constrain the properties that dark matter should have. 

So if $\lambda$ can decay, the change of spacetime energy during the evaporation process consists of two parts, the first is from the thermal Hawking radiation (mainly composed of photons) and the second part is from the emitted dark sector particles.  Thus assuming the thermal radiation does not carry the dark sector particles, the evaporation is governed by the relation 
\begin{equation}\label{first}
    \frac{dE}{dt}=-a b_{c}^{2}T^{4}+\frac{1}{2}\log \frac{r_{halo}}{r_{h}} \frac{d\lambda}{dt}
\end{equation}
which is consistent with the thermodynamic first law. The first term is due to thermal radiation which obeys Stephan-Boltzmann $T^{4}$ law and the second term is due to the decay of dark sector density, there are also the same discussions for RN case in Ref.\cite{Hiscock:1990ex,Xu:2019wak}. From Eq.(\ref{first}),the evolution of horizon radius reads: 
\begin{equation}
    \frac{dr_{h}}{dt}=-a b_{c}^{2} \frac{2r_h}{r_{h}-\lambda} (\frac{r_{h}-\lambda}{4\pi r_{h}^{2}})^{4}.\label{rht}
\end{equation}
Without loss of generality, we can choose Stephan-Boltzmann constant to be $a=1$ from now on. 

To get what $b_{c}$ is, we consider the radiation dominated by massless particles, such as photons. Therefore we apply the geometrical optics approximation and assume all the emitted massless particles move along null geodesics. Using normalized affine parameter $\tau$ we can obtain the geodesic equation
\begin{equation}
	\left(\frac{\mathrm{d} r}{\mathrm{~d} \tau}\right)^{2}=\epsilon^{2}-j^{2} \frac{f(r)}{r^{2}}
\end{equation}
where $ \epsilon=f(r) \frac{\mathrm{d} t}{\mathrm{~d} \tau} $ is the energy and $ j=r^{2} \frac{\mathrm{d} \theta}{\mathrm{d} \tau} $ is the angular momentum of emitted particles. We define a parameter called impact parameter $ b\equiv\frac{j}{\epsilon} $, and the massless particle can reach infinity with condition
\begin{equation}
	\frac{1}{b^{2}}\geqslant V(r)= \frac{f(r)}{r^{2}},
\end{equation}
where $ V(r)=\frac{f(r)}{r^{2}} $ is what we call effective potential.
The maximal value of effective potential gives the critical impact parameter $b_{c}$ which has the equation. 
\begin{equation}
    b_{c}^{2}=\frac{r_{p}^{2}}{f(r_{p})}=\frac{3r_{p}^{3}}{r_{p}-\lambda}
\end{equation}
where $r_{p}$ is given by equation\cite{Gao:2023ltr}
\begin{equation}
  \partial_{r}V(r)|_{r=r_{p}}=  -3r_{h}+2r_{p}+\lambda+3\lambda \log(\frac{r_{h}}{r_{p}})=0.
\end{equation}
$r_{p}$ can be solved in terms of the Lambert W function, which is given by 
\begin{equation}
    r_{p}=-\frac{3}{2} \lambda W_{-1}(-\frac{2 e^{\frac{1}{3}-\frac{r_{h}}{\lambda}}r_{h}}{3\lambda}).
\end{equation}

As we assume that the pair production of the dark sector particle is similar to the QED case, the particle pair production rate per unit volume in four-dimensional spacetime is \footnote{We assume here the horizon radius is much larger than the compton wavelength of the dark sector particle, thus we can approximately use the flat space result for $\Gamma$. } \cite{Hiscock:1990ex,Xu:2019wak}
\begin{equation}
    \Gamma=\frac{2}{(2\pi)^{3}}\left(\frac{m_\lambda}{\hbar}\right)^4\left(\frac{\mathcal{E}}{\mathcal{E}_c}\right)^{2}\exp\biggl\{\frac{-\pi \mathcal{E}_c}{\mathcal{E}}\biggr\}.
\end{equation}
where $m_\lambda$ means the mass of the dark sector particle. Here there exists a critical field strength $\mathcal{E}_{c}$ which depends on the mass and charge of specific dark sector particles. For field strength below the critical value, the production rate is exponentially suppressed, which is why we fail to see the spontaneous generation of dark sector particle-antiparticle pairs under ordinary conditions. 

Analogous to RN black hole case, here $\lambda$ plays the same role as the electric charge. Moreover, in the RN case the chemical potential of charge is the electric potential difference between infinity and horizon, thus here, from the chemical potential of the dark sector, we can deduce that the dark sector potential follows the logarithmic law
\begin{equation}
    \varphi \sim \log r, 
\end{equation}
so the field strength of the dark sector should be
\begin{equation}
    \mathcal{E} \sim \partial_{r} \varphi \sim \frac{1}{r}.
\end{equation}

Note that in order to plot the evolution of $r_{h}$ and $\lambda$, we should redefine the dimensionless quantity by $\frac{r}{r^{0}_{h}} \to r $ and $ \frac{\lambda}{r_{h}^{0}} \to \lambda $, where $r_{h}^{0}$ is the initial black hole horizon radius. Critical field strength $\mathcal{E}_{c}$ for the dark sector may be extremely large, while here, we can still set $m_\lambda=1$ and critical charge $\mathcal{E}_c=1$ (nondimensionalized by $r_{h}^{0}$) as we only focus on qualitative evaporating behavior.
Thus the production rate reads
\begin{equation}
    \Gamma=\frac{e^{-2 \pi  r}}{16 \pi ^3 r^2} \sim O( \frac{1}{r^{2}} \mathrm{Exp}[-r]).
\end{equation}

The dark sector loss rate of the black hole is obtained by integrating the above formula over the entire space outside
the horizon. We have 
\begin{equation}
    \frac{\mathrm{d}\mathcal{\lambda}}{\mathrm{d}t}=-4\pi \int_{r_{h}}^{r_{halo}}\Gamma r^2 \mathrm{d}r=\frac{e^{-2 \pi  r_{halo}}-e^{-2 \pi  r_{h}}}{8 \pi ^3}\simeq \frac{-e^{-2 \pi  r_{h}}}{8 \pi ^3},\label{lt} 
\end{equation}
where we use the condition $r_{halo}\gg r_{h}$ to ignore the $e^{-2 \pi  r_{halo}}$ term in the last step. Since the dark sector potential is of the logarithmic confining type, $\Phi=0$ for $r>r_{halo}$, thus the field strength will vanish when the distance exceeds $r_{halo}$. The pair production rate $\Gamma$ will vanish above $r_{halo}$, so in Eq.(\ref{lt}) we set the upper limit of the integration to be $r_{halo}$. We can also see that because pair production rate $\Gamma$ is exponentially decaying with respect to $r$, almost all the pair production happens near the horizon. This fact is consistent with the second term at the RHS of Eq. (\ref{first})

\begin{figure}[!ht]
	\centering
\subfloat[]
{\includegraphics[height=0.32\textwidth,
width=0.45\textwidth]{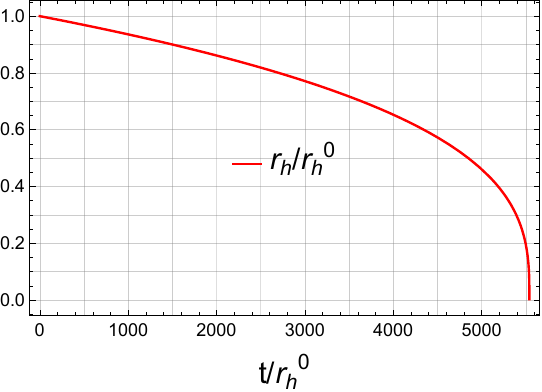}}
\subfloat[]
{\includegraphics[height=0.32\textwidth,width=0.45\textwidth]{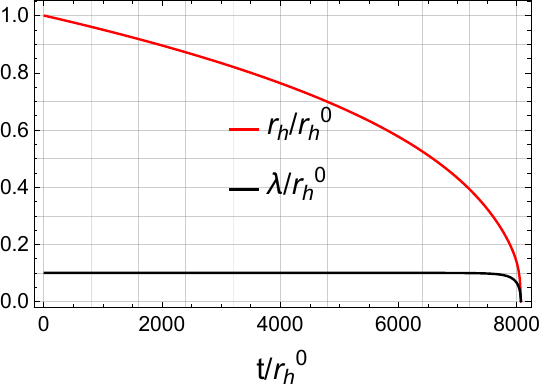}}\\
\subfloat[]
{\includegraphics[height=0.32\textwidth,width=0.45\textwidth]{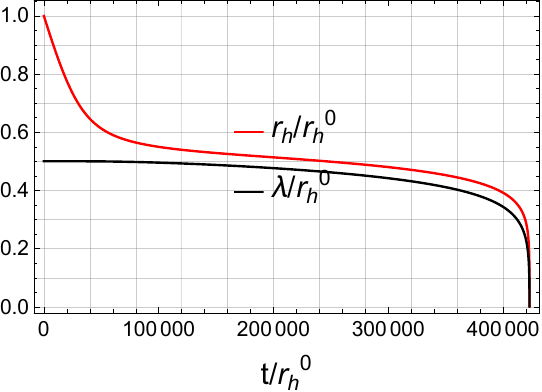} }
\subfloat[]
{\includegraphics[height=0.32\textwidth,width=0.45\textwidth]{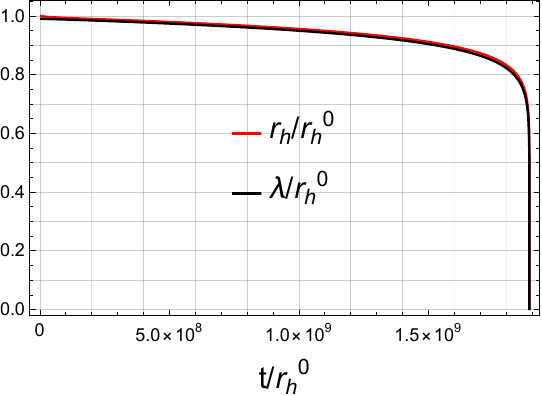}}
 \caption{\textbf{Black hole evaporation process.}The first picture is the Schwarzschild evaporation process. And the next three pictures are PFDM black hole evaporation for different initial ratios $\lambda_{0}/r^{0}_{h}$. The initial ratio $\lambda_{0}/r^{0}_{h}$ are 0.1, 0.5, 0.99 respectively from (b) to (d). We see from the figure that the presence of dark sector $\lambda$ can significantly extend the lifetime of black holes. }
 \label{evap3}
\end{figure}

Solving the two coupled differential equations Eq.(\ref{lt}) and Eq.(\ref{rht}) together, we can get the evolution of dark sector density $\lambda$ and black hole horizon radius $r_{h}$ in Fig.\ref{evap3}. 

We can see from Fig.\ref{evap3} that the initial ratio of dark sector density to the event horizon radius $\lambda_{0}/r^{0}_{h}$ has a great effect on the black hole evaporation process. When the initial ratio is small, which means $r^{0}_{h}\gg \lambda_{0}$, the black hole temperature is relatively high, and $r_{h}$ evaporates quickly to a small value; When the initial ratio is intermediate, even though initial temperature is not zero, due to slow evaporation rate of $\lambda$ in Eq.(\ref{lt}), $r_h$ will evaporate rapidly to $\lambda$ and make temperature close to zero. Then the black hole will experience a long period of slow evaporation as the temperature is nearly zero; when the initial ratio is large, the black hole is nearly extremal initially, thus it will have a longer period of slow evaporation. Thus PFDM black hole  shares the same evaporation behaviors as RN case \cite{Hiscock:1990ex,Xu:2019wak}

Thus in this model, the presence of dark sector field $\lambda$ can extend the lifetime of black holes a lot compared with the Schwarzschild case. The physical process is  similar to the RN black hole case in \cite{Hiscock:1990ex,Xu:2019wak}. While in the real world, the black hole has almost zero charge due to charge neutralization by the interstellar medium, which seems that the enhancement of the black hole's lifetime may not exist in real observation. While here we point out that there still remains the possibility for this lifetime enhancement in the presence of the dark sector.\footnote{Since in this model, the dark sector tends to form halos due to confining potential, it is less possible for the interstellar medium to carry dark sector particles. } Moreover, it is also interesting to see if we could use the observation of black hole lifetime and evaporating behavior to constrain the properties that the dark sector should have. 

\section{Conclusion and Discussion}
In this work, we have investigated the thermodynamics of PFDM black holes. Using the Misner-Sharp mass inside the dark sector halo as the total energy of black hole space-time, we establish the thermodynamic first law. Different from previous work, where mass parameter $M$ is used as the energy of spacetime, our thermodynamic first law uses different definitions of energy. As the mass parameter and thermodynamic first law in Ref.\cite{Cao:2021dcq, Xu:2016ylr} have some ambiguities, for example, the mass parameter does not coincide with the Komar energy of spacetime and the chemical potential of the dark sector relies on the choice of reference point,
our interpretation will help to elucidate these ambiguities. 

After knowing the thermodynamics, we investigate the evaporation behavior of this PFDM black hole. Assuming the vacuum can excite dark sector particle anti-particle pairs, we discuss the effects of the dark sector on black hole evaporation. Like RN black hole case \cite{Hiscock:1990ex,Xu:2019wak}, the presence of a dark sector can also hugely extend the lifetime of a black hole. 

There are various further directions to explore.  Firstly, throughout this paper, the perfect fluid approximation is used. This approximation is so simple that it may hide some important underlying physics. We will try to build the microscopic model of the dark sector to calculate the Schwinger pair production rate and give a direct derivation of dark sector density decay just as the electric charge decreases in RN black hole case.  Secondly, as the dark sector may extend the lifetime of a black hole, it is interesting to see if this mechanism can get support from the observation. 

\begin{acknowledgments}
We are grateful to our group members for useful discussions. This work is supported by the National Natural Science Foundation of China (NSFC) under Grant No.12175105, No.11575083, No.11565017, Ya-Peng Hu is supported by Top-notch Academic Programs Project of Jiangsu Higher Education Institutions (TAPP), Yu-Sen An is supported by Shanghai Post-doctoral Excellence Program No.2021006 and also supported by China Postdoctoral Science Foundation No.2022M710801.
\end{acknowledgments}

\appendix
\section{Relation between Misner Sharp energy and thermodynamic energy }\label{secdil}
In the Schwarzschild black hole, the different energy definitions, such as the energy in the thermodynamic first law, the Komar energy, and the Misner-Sharp energy inside the horizon, all coincide with the mass parameter. But it is not the case when there is a matter field. We illustrate the difference between these energy definitions in this section. 

First, we elucidate this difference in the RN example. 
In stationary black hole spacetime, the apparent horizon coincides with the event horizon, so at $r=r_{h}$, the Misner-Sharp energy inside the horizon is given by $M_{MS}=\frac{r_{h}}{2}$. 
For RN black hole, the blackening factor is 
\begin{equation} \label{rnm}
f(r)=1-\frac{2M}{r}+\frac{Q^{2}}{r^{2}}.
\end{equation}
The outer horizon is solved to be 
\begin{equation}
r_{h}=M+\sqrt{M^{2}-Q^{2}}.
\end{equation}
The Misner Sharp mass inside the horizon radius is $M_{MS}=\frac{1}{2}(M+\sqrt{M^{2}-Q^{2}})$. 

Consider the adiabatic process in thermodynamics, suppose we start from the Schwarzschild black hole with horizon radius $r=r_{h}$, where the thermodynamic first law reads
\begin{equation}
dM_{MS}=TdS.
\end{equation}
By adding charge to this black hole adiabatically (fixing S (or $r_{h}$)), we can get RN black hole with metric (\ref{rnm}). 

Since the chemical potential of electric charge is $\mu=\frac{q}{r_{h}}$, so the work done by adding charge Q is 
\begin{equation}
W=\int_{0}^{Q} \frac{q}{r_{h}}dq=\frac{Q^{2}}{2r_{h}}=\frac{1}{2}\frac{Q^{2}}{M+\sqrt{M^{2}-Q^{2}}}=\frac{1}{2}(M-\sqrt{M^{2}-Q^{2}}).
\end{equation}
As we add charge adiabatically, the horizon radius of the resulting RN black hole is the same as the Schwarzschild black hole before adding charge, so the Misner Sharp energy is the same. Before adding charge, Misner Sharp energy is the total energy. Combining the work done by adding charge gives the total energy of the final RN black hole,
\begin{equation}
W+M_{MS}=M.
\end{equation}
We see the mass parameter M in metric function (\ref{rnm}) is the total energy of spacetime and we can also see the relation between Misner-Sharp energy and total energy. 

%The increased energy by adding charge is given by 

%So the total energy is given by 
%\begin{equation}
   % E=M+\frac{Q^{2}}{4M}=\frac{r_{h}}{2}+\frac{Q^{2}}{2r_{h}}
%\end{equation}
%For RN black hole, if the total energy is E, then from the metric function we can solve 
%\begin{equation}
%r_{h}=E+\sqrt{E^{2}-Q^{2}}
%\end{equation}
The traditional metric function of RN black hole is expressed using Komar mass, when using Misner-Sharp mass, we can see that the metric function is written as 
\begin{equation}
-g_{tt}=1-\frac{2M}{r}+\frac{Q^{2}}{r^{2}}=1-\frac{2M_{MS}}{r}+\frac{Q^{2}}{r^{2}}-\frac{Q^{2}}{2M_{MS}r}.
\end{equation}
The situation is the same in the PFDM case, where the energy of original Schwarzschild spacetime  is $E_{0}=\frac{r_{h}}{2}$. By adding the dark sector adiabatically (fixing $r_{h}$), then the total energy is 
\begin{equation}
E=E_{0}+\mu \lambda=\frac{r_{h}}{2}+\frac{\lambda}{2} \ln \frac{r_{halo}}{r_{h}}.
\end{equation}
To make an analogy with the RN black hole, we should use total energy in the metric instead of using the mass parameter, thus the metric form without confusion is denoted as in Eq. (\ref{metric})

% using $K_{ab}=\nabla_a n_b=\partial_a n_b-\Gamma^c_{ab}n_c$
%\begin{equation}
 %  K_{ab}= \tanh(\rho+\tilde\rho) h_{ab}+\begin{bmatrix}
%\frac{\partial^2\tilde\rho}{\partial t^2}+\frac{1}{y}\frac{\partial\tilde\rho}{\partial y} & \frac{\partial^2\tilde\rho}{\partial t\partial y}+\frac{1}{y}\frac{\partial\tilde\rho}{\partial t} \\
%\frac{\partial^2\tilde\rho}{\partial t\partial y}+\frac{1}{y}\frac{\partial\tilde\rho}{\partial t} & \frac{\partial^2\tilde\rho}{\partial y^2}+\frac{1}{y}\frac{\partial\tilde\rho}{\partial y} \\
%\end{bmatrix},
%\end{equation}
%where $h_{ab}$ is the induced metric on the brane up to order $\mathcal{O}(\frac{\tilde{\rho}^2}{{\rho_0}^2})$.

% The \nocite command causes all entries in a bibliography to be printed out
% whether or not they are actually referenced in the text. This is appropriate
% for the sample file to show the different styles of references, but authors
% most likely will not want to use it.

% The \nocite command causes all entries in a bibliography to be printed out
% whether or not they are actually referenced in the text. This is appropriate
% for the sample file to show the different styles of references, but authors
% most likely will not want to use it.
\nocite{1}

\end{document}